\documentclass[11pt]{article}
\usepackage{moriond,epsfig}

\newcommand\twofigs[5]{
        \begin{center}
	 \begin{figure}[#1]
	  \rule{5cm}{0.2mm}\hfill\rule{5cm}{0.2mm}
           \begin{minipage}{7.2truecm}
            \psfig{file=#2.ps,width=7.1truecm}
            \caption{#3}\label{#2}
           \end{minipage}~\begin{minipage}{7.2truecm}
            \psfig{file=#4.ps,width=7.1truecm}
            \caption{#5}\label{#4}
           \end{minipage}
	  \rule{5cm}{0.2mm}\hfill\rule{5cm}{0.2mm}
         \vskip -.4truecm
	 \end{figure}
        \end{center}
}
\def\NbbD{{N2\beta D}}
\def\eV{\ eV}

\def\MeV{\ MeV}

\def\be{\begin{equation}}
\def\ee{\end{equation}}
\def\bea{\begin{eqnarray}}
\def\eea{\end{eqnarray}}
\def\ba{\begin{array}}
\def\ea{\end{array}}
\begin{document}
\vspace*{4cm}
\title{Phenomenological implication of KamLAND on
 lepton mass matrices}

\author{V.~Antonelli,
 F.~Caravaglios,
 R.~Ferrari,
 \underline{M.~Picariello}}

\address{Dipartimento di Fisica, Universit\`a di Milano
via Celoria 16, I20133 Milano, Italy \\
and INFN, Sezione di Milano}

\maketitle\abstracts{
By using a model independent Monte Carlo approach, we study the possible 
structure of charged and neutral lepton mass matrices, under the assumption  
of an U(2) horizontal symmetry (additional to the usual Standard Model ones) 
involving the light fermion generations.  We assume the most general Majorana 
mass matrix for neutrinos. We update the results of our previous
similar study, by inserting in the analysis the recent KamLAND data,
that contributed to find a final solution to the Solar neutrino problem. 
The introduction of the new experimental data reduce the allowed regions in
the nine dimensional parameter space, and show that our procedure
gives stable solutions.
}

\null
\vspace{-2.2cm}
\null

\raisebox{3.0cm}[0pt][0pt]{
\begin{minipage}[t]{0.9\textwidth}
\begin{center}
\includegraphics*[bb=20 20 252 300, width=3.8cm]{io.ps}
\end{center}
\end{minipage}
}

\null
\vspace{-1.cm}
\null

\section{Introduction}
Neutrino experimental physics has done big steps toward a full understanding 
of this interesting sector of elementary particle physics, also thanks to 
the recent data from Solar, atmospheric and reactor neutrino studies.
These experiments have given a direct confirmation of the fact that the 
electron neutrinos produced in the Sun are converted into muon and tauon 
neutrinos before reaching the terrestrial detectors, in the same way as muon 
neutrinos produced by the cosmic rays are converted
 in something else during their fly in the atmosphere and in the Earth.

A simple and direct way to explain these observations is to assume that
 the mass eigenstates of the neutrinos differ from the eigenstates of
 the weak interaction (flavor eigenstates), as well as it happens in the quark
 sector. This gives rise to a new mixing matrix, similar to the CKM one,
 related to the lepton sector: the so called PMNS mixing matrix.
From the theoretical point of view, 
there is still a lack of solid and unique interpretation of the entries of 
these mixing matrix. 

The aim of this work is to give an indication of which class of 
theoretical models can explain better the experimental results in the lepton
 sector.
We put ourselves in the quite general framework of non Abelian
horizontal symmetries, whose advantages has been deeply discussed in
literature~\cite{nonabelian}. 
This means that we restrict our analysis to the models in which there is an 
additional U(2) horizontal symmetry between the two lighter fermion
generations, as explained in~\cite{NEW}. The consequence of this
assumption is the 
fact that the charged lepton mass matrix has a particular structure, with some 
entries equal to zero. The class of matrices respecting this requirement is 
nevertheless quite general and inside this class we are looking for some 
typical textures characterizing the models which are in better agreement with 
the data.

The method used here for the updated neutrino data, including the
recent results of KamLAND experiment, has been already applied to the
study of the 
experimental implications in the quark sector~\cite{stocchi} and in the lepton 
sector~\cite{NEW} for the data before KamLAND~\cite{24}. Hence we refer the 
interested reader to these papers for a more detailed description of the 
adopted  procedure.
This method is based on the observation that all the theoretical models carry 
an intrinsic incertitude, because they are often completely unpredictive on 
the phases of the entries of the mixing matrix. 
It allows us to discriminate the models on the base of the fine tuning on the 
phases needed to explain the experimental data.
The output of our analysis is an indication on which theoretical model 
needs a smaller fine tuning in the phases to be compatible with the 
experiments.

The work is organized as follows:
 in section~{\bf\ref{sec::neutrinophysics}} we discuss the experimental data
 from neutrino experiments and how we implement these experimental
 constraints;
 in section~{\bf\ref{sec::flavour}} we introduce a model independent
 neutrino mass matrix and then we explain the method used to fit the
 experimental data, we show our results 
 and discuss how they can be used to discriminate between different 
theoretical models.
The section~{\bf\ref{sec::conclusion}} is devoted to the conclusions
 and outlook.
\section{Neutrino physics}\label{sec::neutrinophysics}
During the last seventy years big efforts have been done to understand
 neutrino physics.
In particular a lot of experiments and theoretical discussions have
 created our knowledge about neutrino masses and lepton mixing angles.
For a short review see for instance ref~\cite{NEW,2,3} and references
 therein.
In the first part of this section we analysis all the updated
 experimental data coming from different neutrino experiments.
In the second part, instead, we discuss how we implement the experimental
constraints in our analysis.
\subsection{Experimental data from neutrino experiments}
During the years many experiments investigated the problem 
of neutrino masses and they can be classified in different categories:
the direct kinematical searches (observing mainly the thritium beta
decay), the search for Neutrino-less Double $\beta$ Decay ($\NbbD$),
and the experiments looking for signals of neutrino oscillation. In
this last group of experiments 
different neutrino sources are used: atmospheric and Solar neutrinos
and neutrino beams produced by accelerators or nuclear reactors and
detected at a short distance (short baseline experiments) or very long
distance (long baselines) from the production points. 

The set of data we have considered for most of these different
experiments are the same used in our previous analysis~\cite{NEW}  (to
which we refer the interested reader for a detailed discussion of
these sets of data) or the corresponding updated data for the cases in
which the experimental collaborations published a new data analysis
during the last year. 

The main exception is represented by the data concerning the mixing parameters 
determining the Solar neutrino oscillation. In fact very important
results about these mixing parameters have been obtained in the last
December by KamLAND collaboration~\cite{datiKamLAND,KamLANDtutti}. 
KamLAND~\cite{KamLANDdescription} is a reactor anti-neutrino
experiment using anti-neutrino beams of the energy of a few $\MeV$ and
with a baseline of about 200 Km. This experiment played an essential
role in the final solution of the long 
standing Solar neutrino problem, because it is characterized by the
right value of $L/E$ in order to sound the so called LMA region, that
is the region of the mixing parameters already selected by the Solar
neutrino experiments ($\Delta m^2 \simeq 10^{-5} - 10^{-4}$ $\eV$$^2$).

The first KamLAND results, even if characterized by a statistics that
is still quite poor, have been fundamental because they have given an
independent confirmation of the neutrino oscillation hypothesis for
mixing parameters in the 
typical region of Solar neutrinos and of the fact that the solution of
the Solar neutrino problem is given by the Large Mixing Angle (LMA)
solution. This solution is made possible by the interaction of Solar
neutrinos with matter inside the Sun and the Earth (MSW effect) and is
characterized by a mixing angle large, even if not maximal.
A combined analysis of KamLAND data and of the evidences from all the
Solar neutrino experiments~\cite{KamLANDtutti} gives two distinct
sub-regions still compatible with the data inside the LMA
solution. These regions are usually denoted as HighLMA and LowLMA,
because they are characterized by different values of $\Delta m^2$.
As a matter of fact the LowLMA solution is by far the preferred one
from a statistical point of view. 
Therefore in this analysis we are assuming this as the solution of the
Solar neutrino problem, with the following values of the mixing
parameters: 
$$
\log \left(tan^2 \theta_{sol}\right)  =  -0.36 \pm 0.07\,. \quad
\log \left(\frac{\Delta m_{sol}^2}{\eV^2}\right)  = -4.149 \pm 0.022 \, . 
$$
We did not insert into our analysis the results found by the the
 LSND~\cite{31}  collaboration (signal of $\bar{\nu}_\mu \rightarrow
 \bar{\nu}_e$ oscillation with high value of $\delta m^2$), whose
 explanation would require to introduce a sterile neutrino in addition
 to the usual three generations of the Standard Model or more exotic
 theories.
\subsection{Implementation of the experimental constraints}
\label{sec::implementation}
In our analysis we introduced all the experimental constraints
 on neutrino physics that we have discussed in the previous section,
 with the exception of the LSND results.
We implemented the constraints on the mass differences by assuming
 Gaussian errors in logarithmic scale for 
$\Delta m_{atm}^2$,
 $\sin^22\theta_{atm}$, $\Delta m_{sol}^2$ 
and $\tan^2\theta_{sol}$, where the self explanatory notation refers
 to the variables defined in the previous subsection and in our
 previous analysis~\cite{NEW}.
This approximation, based on the fact that our knowledge of
 the neutrino physics parameters is limited only to the order of 
magnitude, introduces a second order indetermination that can be
 neglected for our purposes. 
To apply the experimental constraints we defined the following set of
observables 
\begin{eqnarray}
{\cal O}_i \in \left\{\log\Delta m_{atm}^2, \log\sin^2 2\theta_{atm},
                      \log\Delta m_{sol}^2, \log\tan^2 \theta_{sol},
                      <m_{\nu}> \right\}
\end{eqnarray}
and we introduced the $\chi^2$ function of the configuration ${\cal
R}$ of the neutrino mass matrix:
\begin{eqnarray}\label{chi2}
\chi^2({\cal R}) = \sum_i \left(
    \frac{{\cal O}_i^{th}({\cal R}) - {\cal O}_i^{exp}}{\sigma_i^{exp}}
                          \right)^2\,.
\end{eqnarray}
In the previous equation, ${\cal O}_i^{th}({\cal R})$ is the value of
 the i$^{th}$ observable calculated for the configuration ${\cal R}$,
 and ${\cal O}_i^{exp}$ and $\sigma_i^{exp}$ are respectively the
 experimental value and its error for the same observable.
We also introduced the upper limits for the neutrino masses 
coming from the direct kinematical searches, but these limits
modify in a negligible way our results and we could 
completely neglect them.
\section{Lepton mass matrices}\label{sec::flavour}
The Standard Model (SM) has a very high predictive power which has been
 tested in the high precision measurements of particle physics.
It allows us to obtain quarks, leptons and bosons masses by using the
 Higgs mechanism and by introducing a set of coupling constants, which
 are also related to the particle masses and the so called mixing
 angles.
While this mechanism is implemented in a natural way and experimentally
 confirmed with very high precision, the coupling constants, and then
 the particle masses and the mixing angles, introduced cannot be
 theoretically determined inside the SM.
An explanation can come by assuming that the SM is a low energy
 effective theory of a more fundamental one, where the full Lagrangian
 contains other fields and the coupling constants are simpler.
In this scenario only the light fields appear in the low energy
 spectrum, while heavy fields decouple.
Due to the freedom in taking the full Lagrangian, one can obtain
 models which give the rich spectrum of the SM at low energy.
However symmetries, simplicity and naturalness induce us to decide for a
 model instead of an other.
A criteria for naturalness is assumed to be how much one has to fine
 tune the parameters introduced in the full Lagrangian to reproduce
 the SM, and our analysis is basically founded on this idea.
To emphasize that, we first introduce a model independent
 parameterization of the Lagrangian of the leptons sector and than
 we discuss the method we used to fit the experimental data within
 our parameterization.
\subsection{A model independent parameterization}\label{sec::model}
In our analysis we did not introduce any light particles additional to the ones
 usually introduced in the SM.
In particular we did not introduce any right partner of the neutrinos.
However we introduce an effective Majorana term for the left handed
 neutrinos.
The low energy kinetic terms in the Lagrangian
% ${\cal L}_{lept}$
of the leptons
% $l^R_i$, $l^L_i$ and $\nu_i$
can be characterized, after the $SU(2)\times U(1)$ spontaneous
symmetry breaking,
% as follow: 
%%
%\newcommand\vect[2]{
%        \left(\begin{array}{c} #1\\#2\end{array}\right)}
%\newcommand\vectb[2]{
%        \left(\begin{array}{ccc} #1&,&#2\end{array}\right)}
%%
%\begin{equation}
% {\cal L}_{lept} =
%     \sum_{i}\left(
%         \bar l^R_i\,{\cal D}\!\!\!\!\slash^R\,l^R_i
%         + \vectb{\bar l^L_i}{\bar \nu_i}\,
%                {\cal D}\!\!\!\!\slash^L\, 
%                \vect{l^L_i}{\nu_i}
%        \right) -
%     \sum_{ij}\left(
%          \bar l^R_i\,M_{ij}\,l^L_j
%         + \bar \nu_i\,m_{ij}\,\nu_j
%        \right)\,.
%\end{equation}
%
with a matrix $M$ that gives the masses of the charged leptons
%, the covariant derivative ${\cal D}^{L,R}$ determine the
%interactions with the electro-weak bosons
and a matrix $m$ giving the neutrino masses.
The two matrices give rise to the so called mixing angles.
In particular, the eingenvalues of $M$ are very well determined by
 the experiments on the charged lepton masses, while $m$ and the
 mixing angles are determined by the neutrino experiments discusses in
 section~{\bf\ref{sec::neutrinophysics}}.
Following the hint given by the theoretical models we parameterize
 $M$ and $m$ by introducing an exponential parameterization of the entries
 of the matrices:
\def\lb{\lambda}
\begin{eqnarray}\label{matrix}
m = m_0
\left(
\ba{ccc}
a_{11}\lb^{v_{11}}  &a_{12}\lb^{v_{12}}  &a_{13}\lb^{v_{13}}\\
a_{21}\lb^{v_{21}}  &a_{22}\lb^{v_{22}}  &a_{23}\lb^{v_{23}}\\
a_{31}\lb^{v_{31}}  &a_{32}\lb^{v_{32}}  &a_{33}\lb^{v_{33}}
\ea
\right) &\quad\quad&
M=m_l\left(
\ba{ccc}
0                         & b_{12}\varepsilon_{1}^{1-p} & 0 \\ 
b_{21}\varepsilon_{1}^{p} & \varepsilon_{2}    & b_{23}\varepsilon_{2}^{r} \\ 
0                         & b_{32}\varepsilon_{2}^{d} & b_{33}
\ea
\right)
\end{eqnarray}
 where we put for convenience $m_0 = 0.06 \eV$, $m_l=m_\tau$ and
$\lambda = 0.2$.
The exponents $v_{ij}$ are real number which give rise to the order
 of magnitude of the entry $i$, $j$.
The coefficients $a_{ij}$ and $b_{ij}$ are complex number of module
one and allow us to introduce generic phase for each entry.
We remember that the matrix $m$ is symmetric due to the Majorana
 nature of the quadratic neutrino terms, i.e. we must taken
 $a_{ij}=a_{ji}$ and $v_{ij}=v_{ji}$.
%
%We notice that the parameterization in our approach differs slightly
% from what one usually finds in literature.
%A simple model implies definite values for the exponents $v_{ij}$,
% but has a negligible predictive power on the coefficients
% $a_{ij}$, which are assumed to be of order one and their incertitudes
% depend on the model.
%In our analyses, to compare different models, we have to fix the
% modules of these coefficients and therefore the model dependent
% uncertainty of these coefficients is transferred into an incertitudes
% on the exponents.
%This means that a given model will have not only an intrinsic
% uncertainty on the phases of the coefficients $a_{ij}$ but also a
% theoretical error on the exponents, which depends on the model itself
% and need to be computed every time.
%Moreover any naive theoretical prediction need to be improved, taking
% into account the renormalization group evolution of the mass and
% mixing angles from the high scale down to the electro-weak scale.
%In this philosophy a given model will not be a point but a ball in the
% nine dimensional parameter space.
%
\subsection{Fitting the experimental data: the method}
\label{sec::fitting}
The goal of this work is to extract the values of the exponents
 $v_{ij}$ from the experimental measurements.
A direct fit of the data is not possible, since the number of free
 parameters in eq.~(\ref{matrix}) is much larger than the number of
 observables, three mass eigenvalues plus the mixing angle parameters.
The main obstacle comes from the coefficients $a_{ij}$
 whose phases are often theoretically unpredicted and are assumed to be
 any number in the range $0-2\pi$ while sometimes they are constraints
 up to 1-10\%.
We treat this uncertainty as a theoretical systematic error.
Namely, we have assigned a flat probability to all the coefficients
 $a_{ij}$ with
\begin{eqnarray}
0<arg(a_{ij})<2\pi\,,&\quad\quad&\left|a_{ij}\right| = 1\,.
\end{eqnarray}
For the theoretical models that unpredicted these phases, our analysis
 apply in a straightforward way.
A model will be more competitive than another, in the sense that it
 needs a smaller fine tuning in the phases $a_{ij}$, if the ball in the parameter space corresponding to this models falls
 in a higher density region.
In the other case a more careful discussion is needed but important
 informations on the fine tuning of the phases can be obtained from
 our results.

The exponents $v$ can take any value: in practice we have chosen
 an interval $-2<v_{ij}<4$.
We choose these limits because for a value of $v$ bigger than 4 the relative
 entry is so close to zero that it is negligible; we have checked that
 no $v$ is smaller that $-2$ even if we enlarge the allowed interval.
For any random choice of the coefficients $a_{ij}$ and the exponents
 $v_{ij}$ we get a numerical matrix for the neutrino sectors.
The diagonalization of this matrix gives us three eigenvalues,
 corresponding to the predicted physical neutrino masses, and a
 numerical unitary matrix which correspond to the MNS mixing matrix.
We have collected a large statistical sample of events.
Each one of these events can be compared with the experimental data 
 through a Monte Carlo by using the $\chi^2$ analysis explained in
 section~{\bf\ref{sec::implementation}}.
An event is accepted with probability given by the exponential weight
 $\chi^2$ defined in eq.~(\ref{chi2}) .

Even if our Monte Carlo approach favors most predictive and accurate
 models, we also emphasize that one should not mistake these results
 with true experimental measurements.
They only give us natural range of values for the exponents $v$.
\subsection{The results}\label{sec::results}
We report in figures the correlation between the $i$ entries for
neutral mass matrix with the $p$-entry for the charged lepton
mass matrix, and the correlation between the $p$, $r$ and $d$ entries
for the charged lepton mass matrix.
We show eight figures corresponding to different pairings of the
exponents.
By looking at the six figure contaning the neutral mass matrix
entries, it is
 evident that there is a symmetry:
 the one under the swapping of the $2^{nd}$ and $3^{rd}$ neutrino.
This symmetry is evident by looking at the symmetry under
 exchange of fig.~{\em\ref{pvs12}} and fig.~{\em\ref{pvs13}}, and
\vskip -2.75truecm
\twofigs{t}{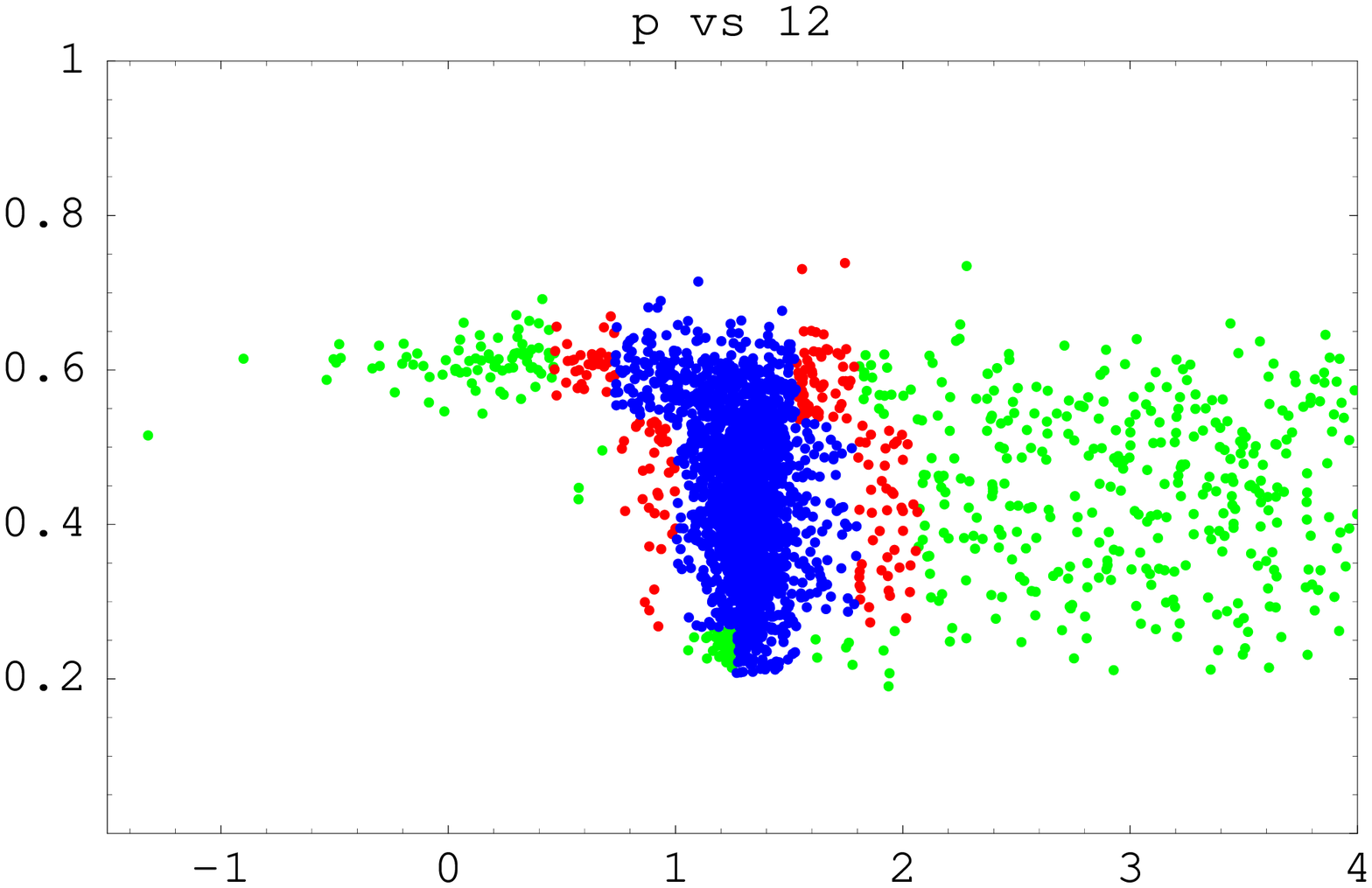}{The exponent $p$ versus the exponent
$v_{12}$.}{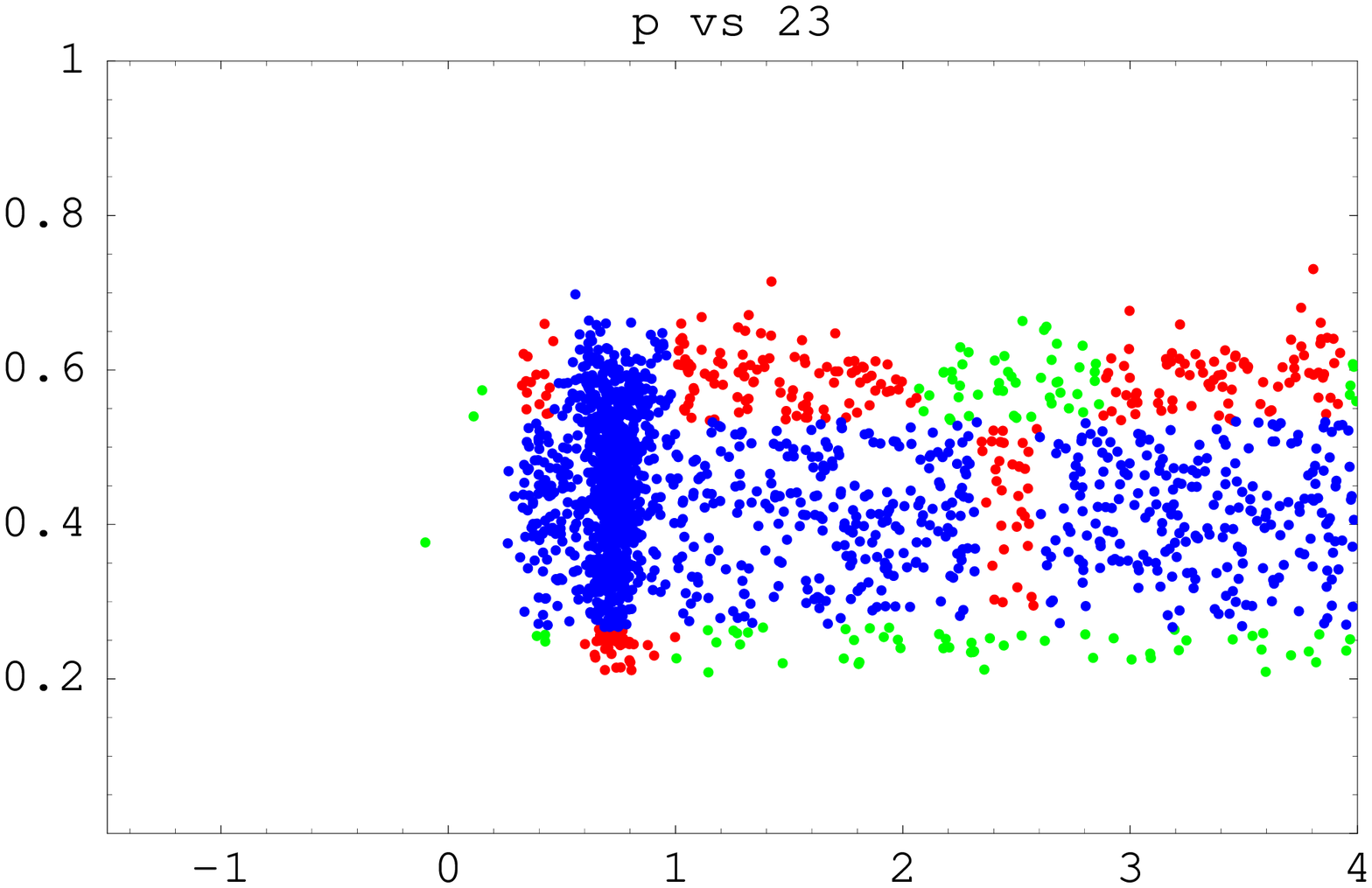}{The exponent $p$ versus the exponent $v_{23}$.}
\twofigs{b}{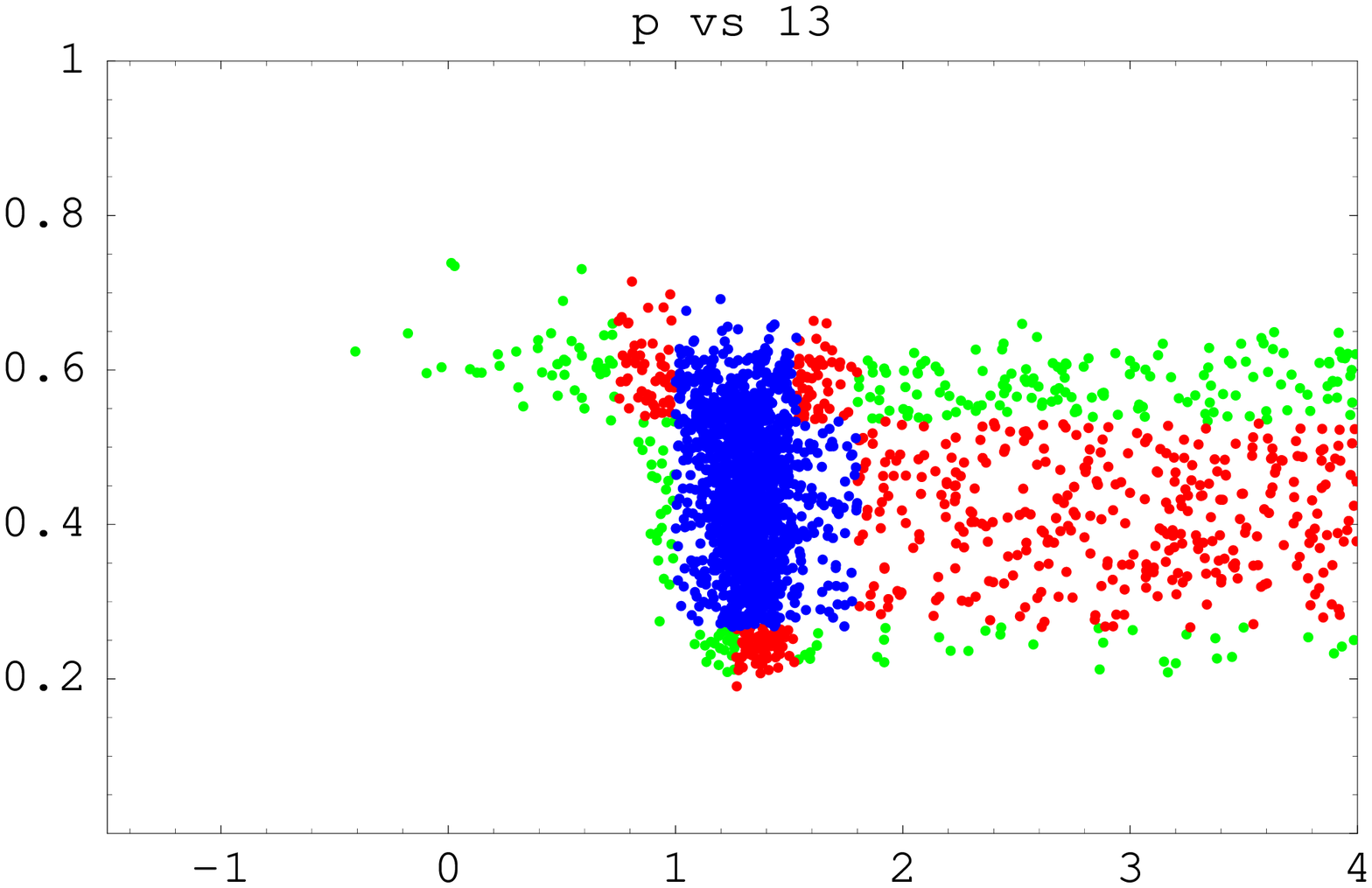}{The exponent $p$ versus the exponent
$v_{13}$.}{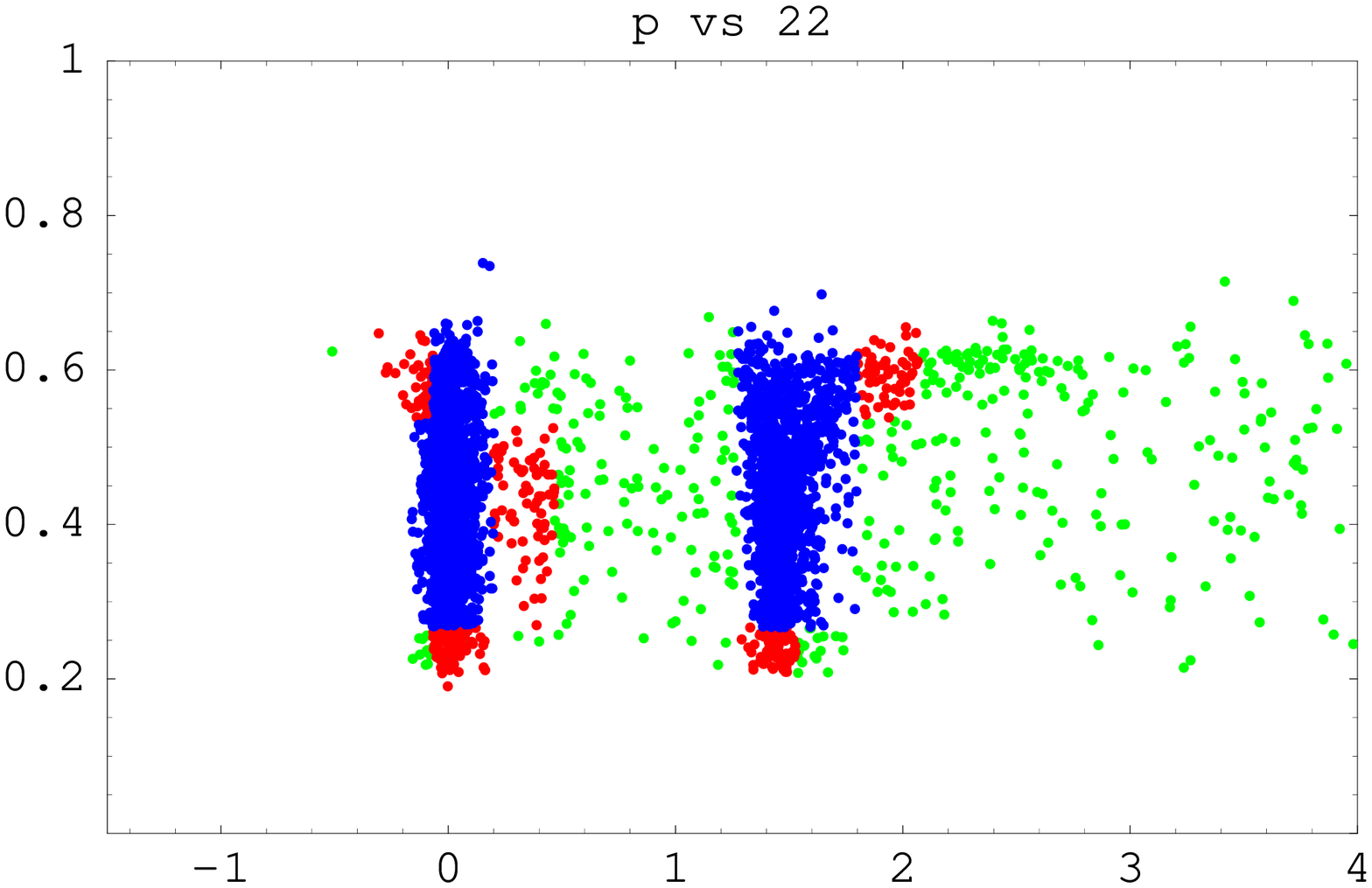}{The exponent $p$ versus the exponent $v_{22}$.}
\twofigs{t}{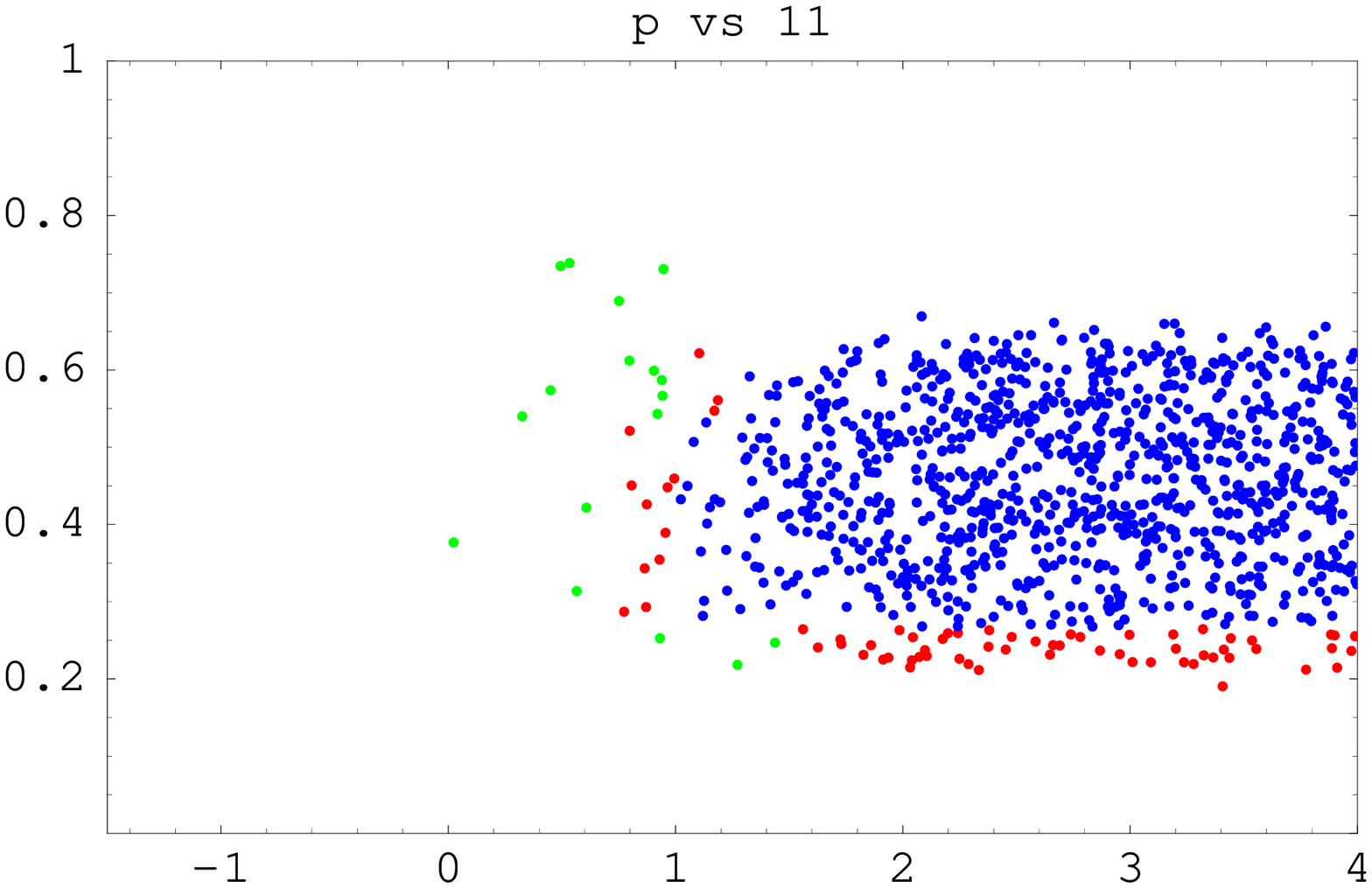}{The exponent $p$ versus the exponent
$v_{11}$.}{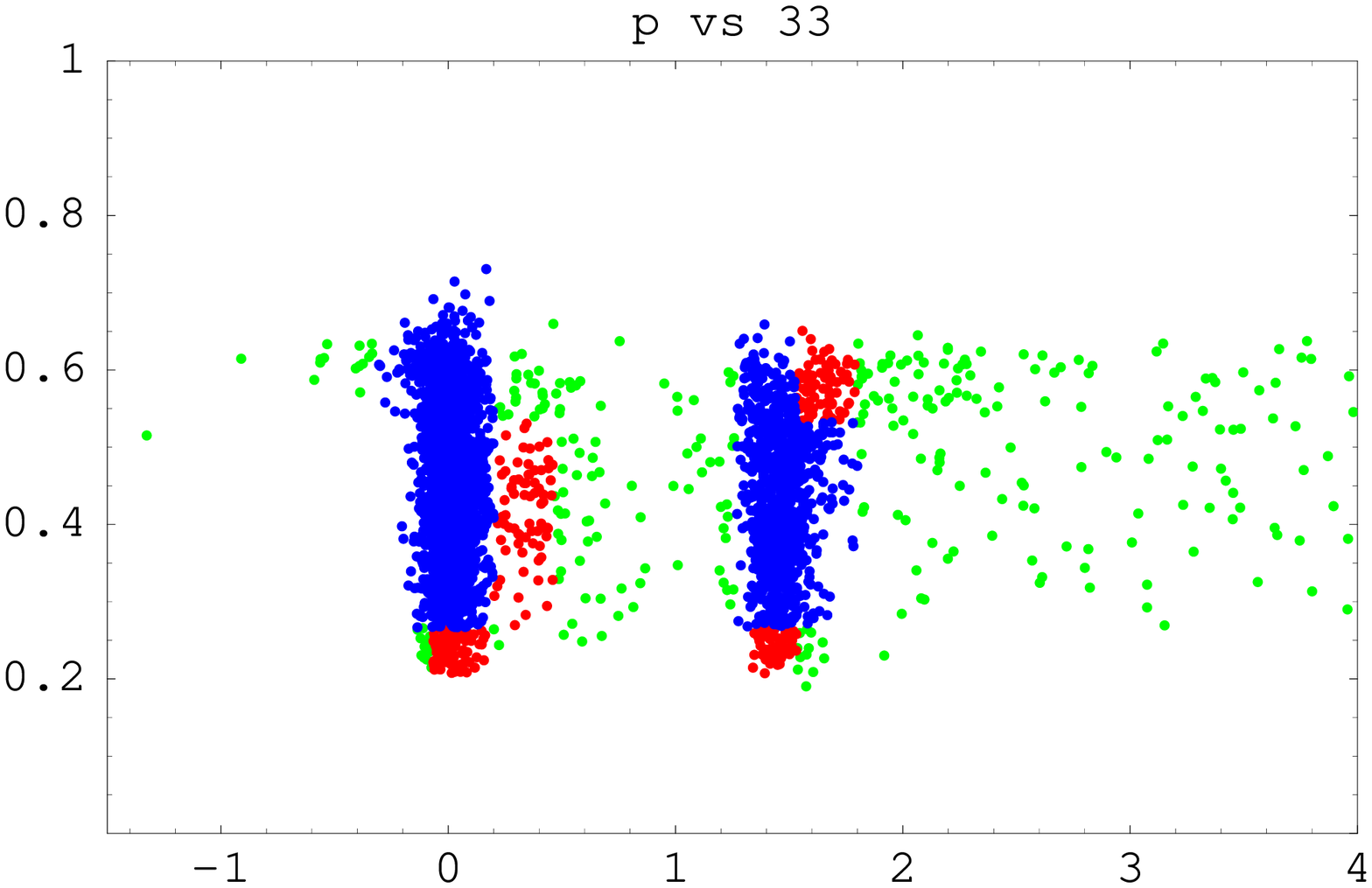}{The exponent $p$ versus the exponent $v_{33}$.}
\twofigs{b}{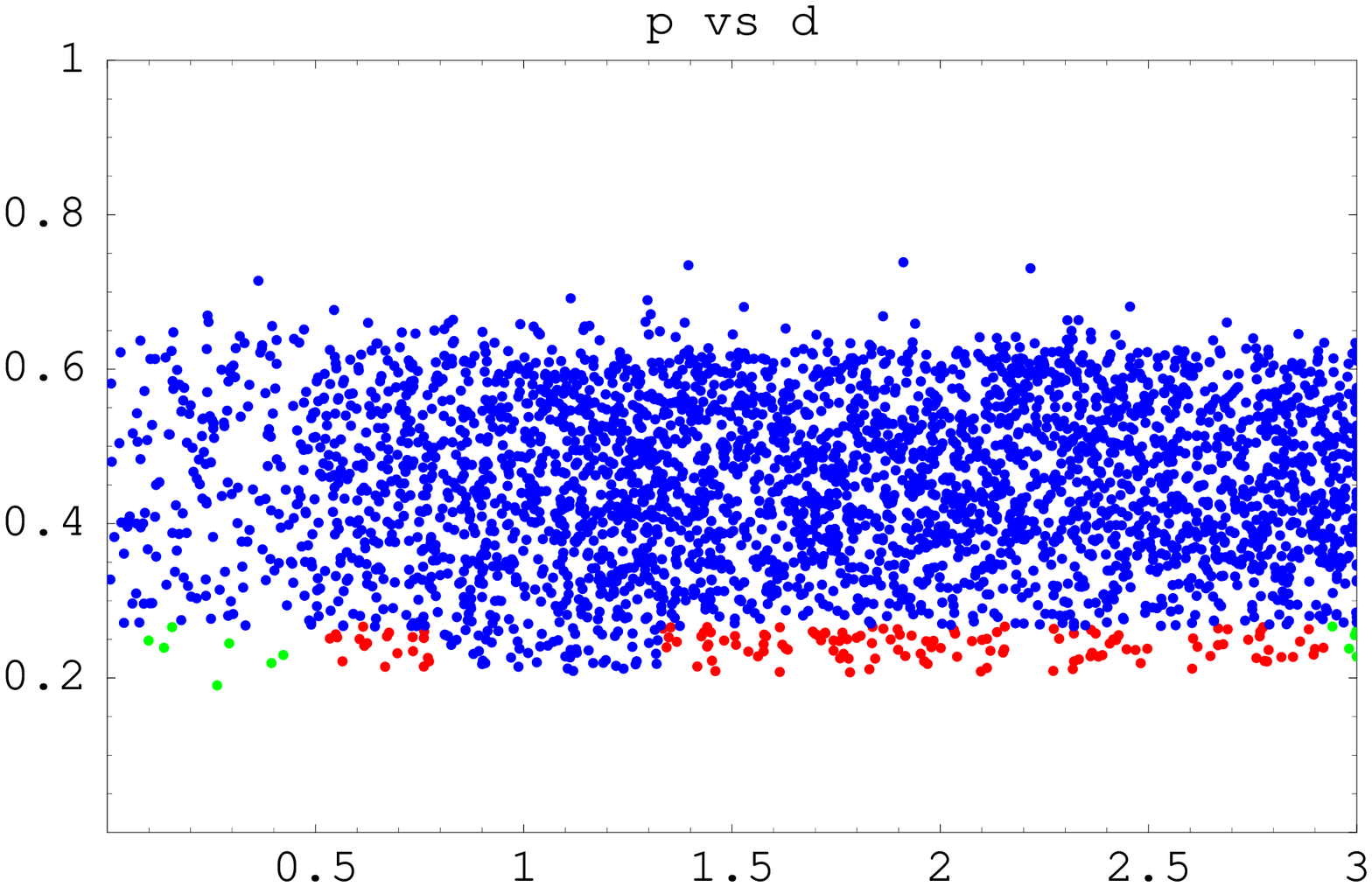}{The exponent $p$ versus the exponent $d$.}{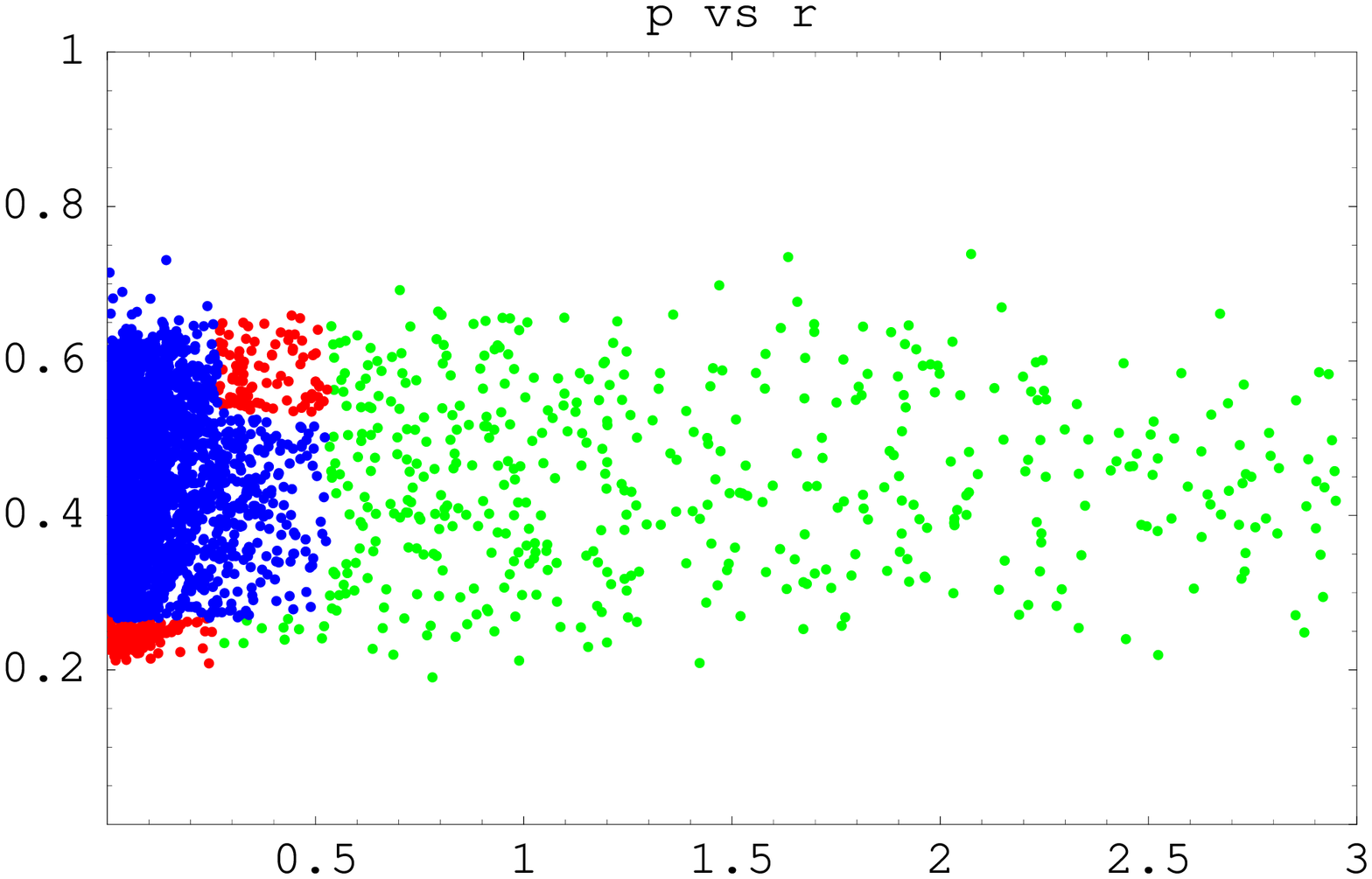}{The
exponent $p$ versus the exponent $r$.}
\vskip -1.15truecm
 the equivalence between fig.~{\em\ref{pvs22}} and fig.~{\em\ref{pvs33}}.
From figs.~{\em\ref{pvs12}}-{\em\ref{pvs13}}-{\em\ref{pvs33}} we see
 that there is the possibility to have the three entries
 $12$-$13$-$33$ very near and in the range $0.06\leftrightarrow0.1 \eV$.
A deeper investigation of the data set gives that the other neutrino
 mass matrix entries are one order of magnitude smaller. 
Moreover the same figures show that the exponent $p$ is about $0.6$.
We notice that this is consistent with the maximal atmospheric
 mixing angle and the large Solar mixing angle, due to the high
 mass degeneracy.
From figures we find the the exponent $p$ is constrained to be between
 $0.2$ and $0.6$.
From fig.~{\em\ref{pvs12}} and fig.~{\em\ref{pvs13}} we find that the
 entries $12$ and $13$ have to be bigger then one except the
 degenerate case considered before, and the case $v_{12}
 (v_{13})\approx 1.2 \pm 0.2$ is prefered.
Fig.~{\em{\ref{pvs23}}} tells us that the entry $23$ is bigger than
 $0.2$ and $v_{23}\approx 0.6\pm 0.3$ is statystically enanched.
Form figs.~{\em{\ref{pvs22}}}-{\em{\ref{pvs33}}} we conclude that the
 exponents $v_{22}$ and $v_{33}$ are bigger than $-0.4$ and the values
$0\pm0.3$ and $1.5\pm0.3$ are the prefered ones.
Finally, form fig.~{\em{\ref{pvs11}}} we obtain that the exponent
 $v_{11}$ is bigger than zero (except in the particular case of {\em
 high} mass degererances).
\section{Conclusions and outlook}\label{sec::conclusion}
Our results show a symmetry under the swap of $2^{nd}$ and $3^{rd}$
 neutrinos: one can interchange the entry $22$ with the $33$,
 and the entry $12$ with the $13$.
We find that there are two kinds of texture which are more compatible
 with the atmospheric, Solar, reactor and kinematic experiments:

$\bullet$
The case where
$$
\ba{cccccc}
v_{11}>1\,;   &v_{12}>1\,;    &v_{13}>1\,;  &v_{22}>-0.3\,;
&v_{23}>0.3\,; &v_{33}>-0.3\,;\\
&0.2<p<0.6\,; &d>0\,;        &r<0.5\,.
\ea
$$
This non-degenerate case have the entries $22$, $23$ and $33$ of order
 $0.06 \eV$ (to explain the large mixing angle and the
 $\Delta m^2_{atm}$), while the other entries have to be at least one order
 of magnitude smaller (to explain the $\Delta m^2_{sol}$).

$\bullet$
The quasi-degenerate case, in which the neutrino masses are
 bigger than $0.06 \eV$ (but smaller than $0.7 \eV$).
This degenerate case can be reached in tree ways:
one diagonal entry and the off-diagonal entry of the residual 2 by 2
matrix block between $0.06$ and $0.7 \eV$, and all the other entries smaller.

In our results there is a hint for an asymmetric non trivial texture
in the charged lepton sector.
Combined with an analogous result in the down quark mass
matrix~\cite{stocchi},
this can be interpreted as a new hint of $SU(5)$ grand
unification~\cite{NEW}.
%In addition to the well know bottom-tau unification, the entries 32 of
%the down quark and charged lepton mass matrices seem to unify. Given
%also that $m_{s}/m_{b}\sim m_{\mu }/m_{\tau }$ we can conclude that
%$SU(5)$ unification works sufficiently well in the $2\times 2$
%sub-matrix formed with the second and third generations. In other
%words, not only the physical mass unify but also the large mixing
%angle between the  right-handed strange and the bottom  unifies with
%the large mixing angle between the left-handed muon and tau
%leptons. The latter is needed to explain the large atmospheric
%neutrino mixing angle and the former can better fit the CKM parameters
%\cite{stocchi}. This left-right asymmetric texture  suggests that the
%\textit{naive } $U(2)$  must be improved: while the 10 of $SU(5)$  can
%succesfully transform as a doublet under $U(2)$,  the same choice for
%the $ \bar 5$ seems disfavoured by data. Taking the $\bar 5$  as
%singlets under $U(2)$, can better accomodate the large left-right
%asymmetry. This choice is also supported by the neutrino data. The
%left-handed neutrino components belongs to $F$, thus if these were
%doublet under $U(2)$ we would expect a relatively large hierarchy
%among different neutrino matrix elements.
%
\section*{Acknowledgments}
We are deeply grateful to E.~Torrente-Lujan and to P.~Aliani for many useful discussions about neutrino physics.
We acknowledge the financial support of the Italian MIUR.
The numerical computations have been performed in the computer farm of
 the Milano University theoretical group.
\section*{References}

\end{document}